\documentclass[sigconf,natbib=true,nonacm]{acmart}
\usepackage{multirow}
\usepackage{float}
\usepackage{subfigure}
\usepackage{enumerate}
\usepackage{enumitem}
\usepackage{stfloats}
\usepackage{hyperref}
\usepackage{breakurl}
\usepackage{diagbox}
\usepackage{balance}
\usepackage{titlesec}
\usepackage{verbatimbox}
\usepackage{indentfirst}
\usepackage[linesnumbered,ruled,vlined]{algorithm2e}

\AtBeginDocument{%
  \providecommand\BibTeX{{%
    \normalfont B\kern-0.5em{\scshape i\kern-0.25em b}\kern-0.8em\TeX}}}
\newcommand{\tabincell}[2]{\begin{tabular}{@{}#1@{}}#2\end{tabular}}
\setcopyright{acmcopyright}
\copyrightyear{2018}
\acmYear{2018}
\acmDOI{10.1145/1122445.1122456}

\acmConference[Woodstock '18]{Woodstock '18: ACM Symposium on Neural
  Gaze Detection}{June 03--05, 2018}{Woodstock, NY}
\acmBooktitle{Woodstock '18: ACM Symposium on Neural Gaze Detection,
  June 03--05, 2018, Woodstock, NY}
\acmPrice{15.00}
\acmISBN{978-1-4503-XXXX-X/18/06}

\DeclareMathOperator\sign{sign}
\DeclareMathOperator\argmin{argmin}

\begin{document}
\fancyhead{}
\title{Towards Low-loss 1-bit Quantization of User-item Representations for Top-K Recommendation}


\author{
Yankai Chen$^1$,
Yifei Zhang$^1$,
Yingxue Zhang$^2$,
Huifeng Guo$^3$,\\
Jingjie Li$^3$,
Ruiming Tang$^3$,
Xiuqiang He$^3$,
Irwin King$^1$
}
 \affiliation{
  \institution{$^1$The Chinese University of Hong Kong, $^2$Huawei Noah’s Ark Lab Canada, $^3$Huawei Noah’s Ark Lab}
  \city{}
  \country{}
}
\email{ {ykchen, yfzhang, king}@cse.cuhk.edu.hk, {yingxue.zhang, huifeng.guo, lijingjie1, tangruiming, hexiuqiang1}@huawei.com }

\begin{abstract}
Due to the promising advantages in space compression and inference acceleration, quantized representation learning for recommender systems has become an emerging research direction recently.
As the target is to embed latent features in the discrete embedding space, developing quantization for user-item representations with a few low-precision integers confronts the challenge of high information loss, thus leading to unsatisfactory performance in Top-K recommendation.  
In this work, we study the problem of representation learning for recommendation with 1-bit quantization.
We propose a model named \textit{{\underline L}ow-{\underline l}oss {\underline Q}uantized {\underline G}raph {\underline C}onvolutional {\underline N}etwork (L$^2$Q-GCN)}.
Different from previous work that plugs quantization as the final encoder of user-item embeddings, L$^2$Q-GCN learns the quantized representations whilst capturing the structural information of user-item interaction graphs at different semantic levels.
This achieves the substantial retention of intermediate interactive information, alleviating the \textit{feature smoothing} issue for ranking caused by numerical quantization.
To further improve the model performance, we also present an advanced solution named L$^2$Q-GCN$_{anl}$ with quantization approximation and annealing training strategy.
We conduct extensive experiments on four benchmarks over Top-K recommendation task.
The experimental results show that, with nearly 9$\times$ representation storage compression, L$^2$Q-GCN$_{anl}$ attains about $90$$\sim$$99\%$ performance recovery compared to the state-of-the-art model.
\end{abstract}




\maketitle
\section{Introduction}

Recommender systems (RSs), as a useful tool to perform personalized information filtering~\cite{covington2016deep,graphsage}, nowadays play a critical role throughout various Web applications, e.g., social networks, E-commerce platforms, and streaming media websites.
Learning vectorized user-item representations (a.k.a. embeddings) for prediction has become the core of modern recommender systems~\cite{lightgcn,cheng2018aspect,neurcf}.
Among existing techniques, graph-based methods, i.e., \textit{Graph Convolutional Networks} (GCNs), due to the ability of capturing high-order relations in user-item interaction topology, well simulate the \textit{collaborative filtering} process and thus produce a remarkable semantic enrichment to the user-item representations~\cite{ngcf,sun2019multi,graphsage,lightgcn}.
 
Apart from the representation informativeness, space overhead is another important criterion for realistic recommender systems.
With the explosive growth of interactive data encoded in the graph form, quantized representation learning recently provides an alternative option to GCN-based recommender methods for optimizing the model scalability.
Generally, quantization is the process of converting a continuous range of vectorized values into a finite discrete set, e.g., integers.
Instead of using continuous embeddings, e.g., 32-bit floating points, 1-bit quantization however embeds user-item latent features into the binary embedding space, e.g., $\{-1,1\}^{d}$.
By enabling the usage of low-precision integer arithmetics, 1-bit quantized representations have the promising potential in space compression and inference acceleration for recommendation~\cite{tailor2020degree}.

\begin{figure}[tp]
\begin{minipage}{0.5\textwidth}
\includegraphics[width=3.33in]{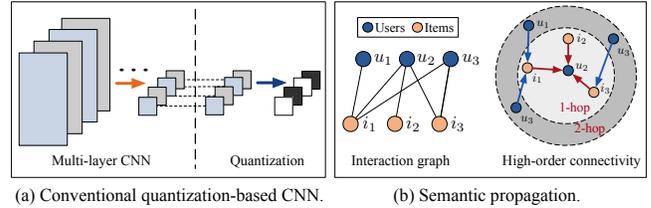}
\end{minipage} 
\setlength{\abovecaptionskip}{0.2cm}
\setlength{\belowcaptionskip}{0.2cm}
\caption{Illustration of conventional quantized CNN and semantic propagation within the user-item interaction graph.}
\label{fig:cnn}
\end{figure}

Despite the previous attempts to quantize traditional model-based RS methods~\cite{zhang2016discrete,zhang2017discrete,kang2021learning}, quantizing graph-based RS models~\cite{hashgnn} however receive less attention so far.
Due to their unique message passing mechanism~\cite{graphsage,kipf2016semi} in encoding high-order interactive information, it is emerging as a good research topic to study representation quantization for GCN-based recommender models.
Currently, it is still challenging to approach this target as previous work falls short of satisfaction in terms of recommendation accuracy.
The crux of this phenomenon is mainly twofold:
\begin{itemize}[leftmargin=*]
\item Intuitively, the performance degradation for quantizing representations is mainly caused by the limited expressivity of discrete embeddings. 
Different from applications in other domains, e.g., Natural Language Processing, Computer Vision~\cite{bennett1948spectra,gersho2012vector,bai2020binarybert}, the top principle of quantization for recommendation is ranking preserving.
However, compared to full-precision embeddings, the vectorized latent features of both users and items tend to be smoothed by the discreteness of quantization naturally.
For instance, after the quantization into binary embedding space $\{-1,1\}^{d}$, only the digit signs are kept, no matter what specific values of continuous embeddings originally are.
Consequently, this leads to the information loss when estimating users' preferences towards different items, thus drawing a conspicuous performance decay in ranking tasks, e.g., Top-K recommendation.  

\item Technically, previous work mainly takes inspiration from the methodology of \textit{Quantized Convolutional Neural Networks}~\cite{qin2020forward,rastegari2016xnor,lin2017towards} (illustrated in Figure~\ref{fig:cnn}(a)) to plug quantization as a separate encoder that is posterior to the GCN architecture. 
Being the final encoder for object embeddings (i.e., users and items), this however ignores the intermediate representations when \textit{graph convolution} performs on the sub-structures of interaction graphs. 
As pointed out by~\cite{ngcf}, the intermediate information at different layers of graph convolution is important to reveal different semantics of user-item interactions.
For example, as shown in Figure~\ref{fig:cnn}(b), when lower-layers propagate information between users and items that have historical interactions, higher-layers capture higher-order proximity of users (or items).
Hence, ignoring intermediate semantics fails to the feature enrichment for embedding quantization. 
Furthermore, due to the ``double-edged'' effect of GCN, final embeddings at the last graph convolution layer may be \textit{over-smoothed~\cite{li2018deeper,li2019deepgcns}} to become uninformative accordingly. 
This implies that simply using the output embeddings may be risky and problematic~\cite{lightgcn}, which leads to the suboptimal quantized representations for recommendation.
\end{itemize}

In this paper, we investigate the problem of 1-bit quantized representation learning for recommendation with the GCN framework.
We propose a model named \textit{ {\underline L}ow-{\underline l}oss 1-bit {\underline Q}uantized {\underline G}raph {\underline C}onvolutional {\underline N}etwork (L$^2$Q-GCN)}.
L$^2$Q-GCN interprets the user-item representations in the discrete space $\{-1,1\}^{d}$.
We design the quantization-based graph convolution such that L$^2$Q-GCN achieves the embedding quantization whilst capturing different levels of interactive semantics in exploring the user-item interaction graphs.
Intuitively, such topology-aware quantization makes the user-item representations more comprehensive, and thus significantly alleviates the information loss of numerical quantization that causes the performance decay.
Specifically, we propose two solutions, namely L$^2$Q-GCN$_{end}$ and L$^2$Q-GCN$_{anl}$, to provide flexibility towards different deployment scenarios.
We conduct extensive experiments on four benchmarks over Top-K recommendation task. To summarize, our main contributions are as follows:
\begin{enumerate}[leftmargin=*]
\item We implement our proposed network design in L$^2$Q-GCN$_{end}$ that is trained in an end-to-end manner. 
Our experimental results demonstrate that, L$^2$Q-GCN$_{end}$ achieves nearly 11$\times$ representation compression and about 40\% inference acceleration, while retaining over $80$\% recommendation capacity, compared to the state-of-the-art full-precision model.

\item To further improve the recommendation performance, we propose an advanced solution L$^2$Q-GCN$_{anl}$ with approximation that is trained by a two-step annealing training strategy. 
With slightly additional space cost, L$^2$Q-GCN$_{anl}$ can attain $90$$\sim$$99\%$ performance recovery. 

\item We release codes and datasets to researchers via the link~\cite{our_code} for reproducing and validating.
\end{enumerate}

\textbf{Organization.} 
We first review the related work in Section~\ref{sec:work} and present L$^2$Q-GCN$_{end}$ and L$^2$Q-GCN$_{anl}$ in Sections~\ref{sec:end} and~\ref{sec:anl}. 
We report the experimental results on four benchmarks in Section~\ref{sec:exp} and conclude the paper in Section~\ref{sec:con}.

\section{Related Work}
\label{sec:work}

\subsection{Full-precision GCN-based RS Models}
Recently, GCN-based recommender systems have become new state-of-the-art methods for Top-K recommendation~\cite{lightgcn,ngcf}, thanks to their capability of capturing semantic relations and topological structures for user-item interactions~\cite{kipf2016semi,graphsage,zhang2019bayesian}.
Motivated by the advantage of graph convolution, prior work such as GC-MC~\cite{gcmc}, PinSage~\cite{pinsage} and recent state-of-the-art models NGCF~\cite{ngcf} and LightGCN~\cite{lightgcn} are proposed. 
Generally, they adapt the GCN framework to simulate the collaborative filtering process in high-order graph neighbors for recommendation.
For example, NGCF~\cite{ngcf} follows the GCN-based information propagation rule to learn embeddings: feature transformation, neighborhood aggregation, and nonlinear activation. 
LightGCN~\cite{lightgcn} further simplifies the graph convolution by retaining the most essential GCN components to achieve further improved recommendation performance. 
In our experiments, we settle these two state-of-the-art full-precision methods as the benchmarking reference for L$^2$Q-GCN in Top-K recommendation.

\subsection{General Binarized GCN Frameworks}
Network binarization~\cite{hubara2016binarized} aims to binarize all parameters and activations in neural networks so that they can even be trained by logical units of CPUs. 
Binarized models will dramatically reduce the memory usage.
Despite the progress of binarization CNNs~\cite{rastegari2016xnor,qin2020forward} for multimedia retrieval, this technique is not adequately studied in geometric deep learning~\cite{bigcn,bahri2021binary,wu2021hashing}.
Bi-GCN~\cite{bigcn} and BGCN~\cite{bahri2021binary} are two recent trials.
However, they are mainly designed for geometric classification tasks, but their capability of link prediction (a geometric form of recommendation) is unclear. 
Furthermore, compared to focusing on the quantization for user-item representations only, a complete network binarization will further abate the numerical expressivity of modeling ranking information for recommendation.
This implies that a direct adaptation of binary GCN models may draw large performance decay in Top-K recommendation. 

\subsection{Quantization-based RS Models}
Quantization-based recommender models are attracting growing attention recently~\cite{hashgnn,kang2021learning,shi2020compositional,xu2020multi}. 
Compared to network binarization, they do not pursue extreme model compression, but focus on quantization for user-item representations with a few integers.

These models can be generally categorized into model-based~\cite{kang2021learning,zhang2016discrete,xu2020multi,zhang2017discrete} and graph-based~\cite{hashgnn}.
HashGNN~\cite{hashgnn}, as the state-of-the-art graph-based solution, takes both advantages of graph convolution and embedding quantization. 
Specifically, HashGNN~\cite{hashgnn} designs a two-step quantization framework by combining GraphSage~\cite{graphsage} and \textit{learn to hash} methodology~\cite{lsh,wang2017survey,erin2015deep,zhu2016deep}.
Generally, \textit{Learn to hash} aims to learn hash functions for generating discriminative codes. 
HashGNN first invokes the two-layer GraphSage as the encoder to get the embeddings for users and items; then it stacks a hash layer to get the corresponding binary encodings afterwards. 
However, the main inadequacy of HashGNN is that, the quantization process only proceeds at the end of multi-layer graph convolution, i.e., using the aggregated output of two-layer GraphSage for representation binarization. 
While multi-layer convolution helps to aggregate local information that lives in the consecutive graph hops, HashGNN may thus not be able to capture intermediate semantics from nodes' different layers of receptive fields, producing a suboptimal quantization of node embeddings. 
Compared to HashGNN, our proposed L$^2$Q-GCN model conducts the quantization-based graph convolution when exploring the user-item interaction graph in a layer-wise manner. We justify the effectiveness in Section~\ref{sec:exp}.

\section{L$^2$Q-GCN Methodology}
\label{sec:end}
\subsection{Problem Formulation}

User-item interactions can be represented by a bipartite graph, i.e., $\mathcal{G} = \{(u, i)|u$ $\in$ $\mathcal{U}$, $i$ $\in$ $\mathcal{I}\}$. $\mathcal{U}$ and $\mathcal{I}$ denote the sets of users and items.
We denote $y_{u,i} = 1$ to indicate there is an observed interaction between $u$ and $i$, e,g., \textit{browse}, \textit{click}, or \textit{purchase}, otherwise $y_{u,i}$ $=$ $0$.

\textbf{Notations.} We use bold lowercase, bold uppercase, and calligraphy characters to denote vectors, matrices, and sets, respectively. Non-bold characters are used to denote graph nodes or scalars. 
Due to the page limit, we summarize all key notations in Appendix~\ref{sec:notation}.

\textbf{Task Description.} Given an interaction graph, the problem studied in this paper is to learn quantized representations $\mathcal{Q}_u$ and $\mathcal{Q}_i$ for user $u$ and item $i$, such that the online recommender system model can predict the probability $\hat{y}_{u,i}$ that user $u$ may adopt item $i$.

\begin{figure*}[tp]
\begin{minipage}{1\textwidth}
\includegraphics[width=7in]{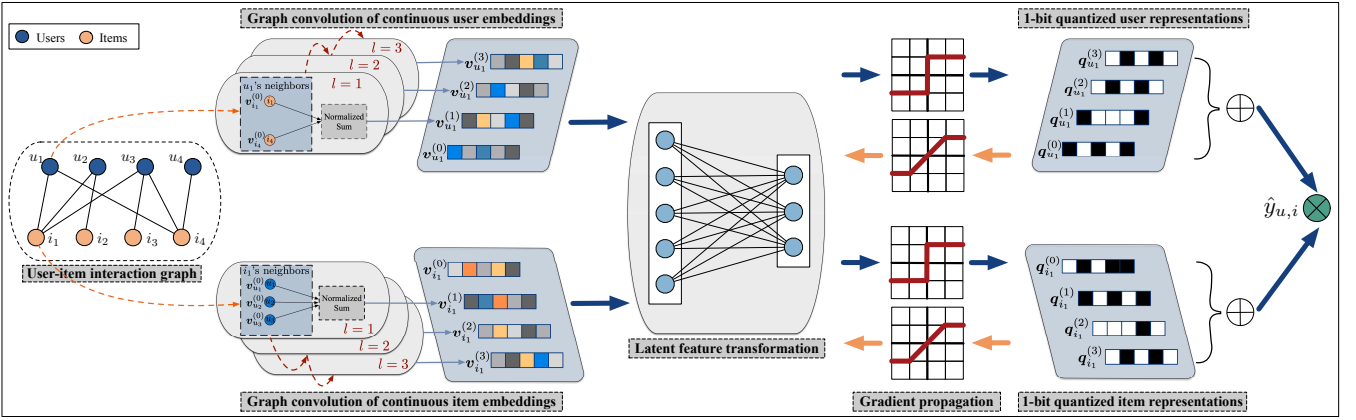}
\end{minipage} 
\setlength{\abovecaptionskip}{0.2cm}
\setlength{\belowcaptionskip}{0.2cm}
\caption{Illustration of L$^2$Q-GCN framwork (Best view in color).}
\label{fig:model}
\end{figure*}

\subsection{Basic Implementation: L$^2$Q-GCN$_{end}$}
The general idea of L$^2$Q-GCN is to learn node representations by propagating latent features via the graph topology~\cite{wu2019simplifying,lightgcn,kipf2016semi}. 
It performs iterative \textit{graph convolution}, i.e., propagating and aggregating information of neighbors to update representations (embeddings) of target nodes, which can be formulated as follows:
\begin{equation}
\setlength\abovedisplayskip{2pt}
\setlength\belowdisplayskip{2pt}
\boldsymbol{v}_x^{(l)} = AGG\left(\{\boldsymbol{v}_y^{(l-1)}: y \in \mathcal{N}(x)\}\right),
\end{equation}
where $\boldsymbol{v}_x^{(l)}$ denotes node $x$'s embedding after $l$ layers of information propagation.
 $\mathcal{N}(x)$ represents $x$'s neighbor set and $AGG$ is the aggregation function aiming to transform the center node feature and the neighbor features.
 We illustrate the framework in Figure~\ref{fig:model}.

\subsubsection{\textbf{Quantization-based Graph Convolution.}}
We adopt the graph convolution paradigm working on the continuous space from~\cite{lightgcn} that recently shows good performance under recommendation scenarios.
Let $\boldsymbol{v}^{(l)}_u\in \mathbb{R}^{c}$ and $\boldsymbol{v}^{(l)}_i\in \mathbb{R}^{c}$ denote the continuous feature embeddings of user $u$ and item $i$ computed in the $l$-th layer.
They can be respectively updated by utilizing information from the ($l-1$)-th layer in iteration as follows:
\begin{equation}
\setlength\abovedisplayskip{2pt}
\setlength\belowdisplayskip{2pt}
\resizebox{1\linewidth}{!}{$
\displaystyle
\boldsymbol{v}^{(l)}_u = \sum_{i\in \mathcal{N}(u)} \frac{1}{\sqrt{|\mathcal{N}(u)|\cdot|\mathcal{N}(i)|}}\boldsymbol{v}^{(l-1)}_i, \ \ \boldsymbol{v}^{(l)}_i = \sum_{u\in \mathcal{N}(i)} \frac{1}{\sqrt{|\mathcal{N}(i)|\cdot|\mathcal{N}(u)|}}\boldsymbol{v}^{(l-1)}_u.
$}
\end{equation}
After getting the intermediate embeddings, e.g., $\boldsymbol{v}^{(l)}_u$, we conduct 1-bit quantization as:
\begin{equation}
\boldsymbol{q}_u^{(l)} = \sign\big(\boldsymbol{W}^T\boldsymbol{v}^{(l)}_u\big), \ \  \boldsymbol{q}_i^{(l)} = \sign\big(\boldsymbol{W}^T\boldsymbol{v}^{(l)}_i\big),
\end{equation}
where $\boldsymbol{W} \in R^{c\times d}$ is a matrix that transforms $\boldsymbol{v}^{(l)}_u$ to $d$-dimensional latent space for 1-bit quantization. 
Function $\sign(\cdot)$ maps floating-point inputs into the discrete binary space, e.g., $\{-1,1\}^d$.
By doing so, we can obtain the quantized embedding {\small$\boldsymbol{q}_u^{(l)}$} whilst retaining the latent user features directly from $\boldsymbol{v}^{(l)}_u$.
For ease of L$^2$Q-GCN$_{end}$'s binarization storage, we can further conduct a numerical translation to these quantized embeddings from $\{-1,1\}^d$ to $\{0,1\}^d$.
After $L$ layers of feature propagation and quantization, we have built the targeted quantized representations $\mathcal{Q}_u$ and $\mathcal{Q}_i$ as:
\begin{equation}
\mathcal{Q}_u = \{\boldsymbol{q}^{(0)}_u, \boldsymbol{q}^{(1)}_u, \cdots, \boldsymbol{q}^{(L)}_u\}, \ \ \mathcal{Q}_i = \{\boldsymbol{q}^{(0)}_i, \boldsymbol{q}^{(1)}_i, \cdots, \boldsymbol{q}^{(L)}_i\}. 
\end{equation}
Both $\mathcal{Q}_u$ and $\mathcal{Q}_i$ track the intermediate information binarized from full-precision embeddings at different layers.  
Intuitively, they represent the interactive information that is propagated back and forth between users and items within these layers, simulating the \textit{collaborative filtering} effect to exhibit in the quantized encodings for recommendation. 

\subsubsection{\textbf{Model Prediction.}}
Based on the quantized representations of users and items, i.e., $\mathcal{Q}_u$ and $\mathcal{Q}_i$, we predict the matching scores by naturally adopting the inner product as:
\begin{equation}
\label{eq:score}
\hat{y}_{u,i} =  f(\mathcal{Q}_u)^T \cdot f(\mathcal{Q}_i),
\end{equation}
where function $f$ abstracts the way to utilize $\mathcal{Q}_u$ and $\mathcal{Q}_i$.
In this paper, we implement $f$ by taking an element-wise summation:  
\begin{equation}
\label{eq:useQ}
f(\mathcal{Q}_u) = \sum_{l=0}^L \boldsymbol{q}^{(l)}_u, \ \ f(\mathcal{Q}_i) = \sum_{l=0}^L \boldsymbol{q}^{(l)}_i. 
\end{equation}

\noindent\textit{\underline{Clarification.} 
In this work, we interpret quantized representations as integers and apply integer arithmetics in computation. 
As we will show later in experiments, this introduces about 40\% of computation acceleration. Certainly, it can be extended to binary arithmetic logics for further optimization, and we leave it for future work.}

\subsubsection{\textbf{Model Optimization.}}
We then introduce our objective and backward-propagation strategy for optimization.

\textbf{Objective Function.}
Our objective function consists of two components, i.e., graph reconstruction loss $\mathcal{L}_{rec}$ and BPR loss $\mathcal{L}_{bpr}$. 
The motivation of such design is basically twofold: 
\begin{itemize}[leftmargin=*]
\item $\mathcal{L}_{rec}$ reconstructs the observed topology of interaction graphs;
\item $\mathcal{L}_{bpr}$ learns the relative rankings of user preferences towards different items. 
\end{itemize}
Concretely, we implement $\mathcal{L}_{rec}$ with the cross-entropy loss: 
\begin{equation}
\resizebox{1\linewidth}{!}{$
\displaystyle
\mathcal{L}_{rec} = \sum_{u \in \mathcal{U}} \sum_{i\in \mathcal{I}} y_{u,i}\ln\sigma\Big((\boldsymbol{v}^{(L)}_u)^T \cdot \boldsymbol{v}^{(L)}_i\Big) + (1-y_{u,i})\ln\Big(1-\sigma\big((\boldsymbol{v}^{(L)}_u)^T \cdot \boldsymbol{v}^{(L)}_i\big)\Big),
$}
\end{equation}
where $\sigma$ is the activation function, e.g., Sigmoid.
$\mathcal{L}_{rec}$ bases on the full-precision embeddings at the last layer, e.g., {\small$\boldsymbol{v}^{(L)}_u$}, providing the latest intermediate information for topology reconstruction as much as possible.
As for $\mathcal{L}_{bpr}$, we employ \textit{Bayesian Personalized Ranking} (BPR) loss~\cite{rendle2012bpr} as follows:
\begin{equation}
\label{eq:hd-bpr}
\mathcal{L}_{bpr} = -\sum_{u \in \mathcal{U}} \sum_{i\in \mathcal{N}(u)} \sum_{j\notin \mathcal{N}(u)} \ln \sigma(\hat{y}_{u,i} - \hat{y}_{u,j}).
\end{equation}
$\mathcal{L}_{bpr}$ relies on the quantized node representations, e.g., $\mathcal{Q}_u$, to encourage the prediction of an observed interaction to be higher than its unobserved counterparts~\cite{lightgcn}.
Finally, our final objective function is defined as:
\begin{equation}
\mathcal{L} = \mathcal{L}_{rec} + \mathcal{L}_{bpr} + \lambda ||\Theta||_2^2, 
\end{equation}
where $\Theta$ is the set of trainable parameters and embeddings, and $||\Theta||_2^2$ is the $L$2-regularizer parameterized by $\lambda$ to avoid over-fitting.

\noindent\textit{\underline{Clarification.} 
Please notice that at the training stage, we usually take the mean value of $f(\mathcal{Q}_u)$ by shrinking it as $f(\mathcal{Q}_u) = f(\mathcal{Q}_u)/(L+1)$ (likewise for $f(\mathcal{Q}_i)$).
Essentially, this useful training strategy~\cite{lightgcn,kgcn} reduces the absolute value of predicted scores to a smaller scale, which significantly stabilizes the training process to avoid the undesirable divergence in embedding optimization; but most importantly, it has no effect on the relative rankings of all scores.}

\textbf{Backward-propagation Strategy.}
Unfortunately, the sign function is not differentiable. This means that the original derivative of the sign function is 0 almost everywhere, making the intermediate gradients accumulated before quantization zeroed here.
To avoid this and approximate the gradients for backward propagation, we adopt the \textit{Straight-Through Estimator}~\cite{bengio2013estimating} with \textit{gradient clipping} as:
\begin{equation}
\left\{ 
\begin{aligned}
& \boldsymbol{q}^{(l)}_u = \sign\big(\phi \big),  &\text{forward propagation} \\
& \frac{\partial \boldsymbol{q}^{(l)}_u}{\partial \phi} := \boldsymbol{1}_{|\phi| \leq 1}. & \text{backward propagation}
\end{aligned}
\right.
\end{equation}
Derivative $\boldsymbol{1}_{|\phi| \leq 1}$ can be viewed as passing gradients e.g., $\Delta$, through \textit{hard-tanh} function~\cite{courbariaux2016binarized}, i.e., $\max(-1, \min(1, \Delta))$.
This passes gradients backwards unchanged when the input of sign function, e.g., $\phi$, is within range of \{-1, 1\}, and cancels the gradient flow otherwise~\cite{alizadeh2018empirical}. We illustrate the process of gradient propagation in Figure~\ref{fig:model}.

So far, we have already introduced the skeleton of our proposed network that learns the quantized representations $\mathcal{Q}_u$ and $\mathcal{Q}_i$ via exploring the interaction graph topology. Since it can be trained in an end-to-end manner, we directly name it as L$^2$Q-GCN$_{end}$.

\begin{figure}[tp]
\begin{minipage}{0.5\textwidth}
\includegraphics[width=3.3in]{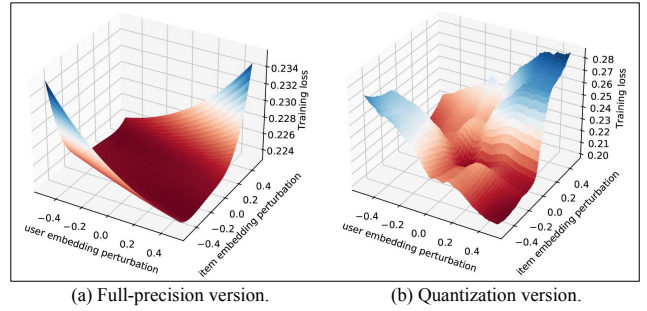}
\end{minipage} 
\setlength{\abovecaptionskip}{0.2cm}
\setlength{\belowcaptionskip}{0.2cm}
\caption{Loss landscapes visualization.}
\label{fig:ste_loss}
\end{figure}

\section{Advanced Solution: L$^2$Q-GCN$_{anl}$}
\label{sec:anl}
Although L$^2$Q-GCN$_{end}$ can achieve the quantization for user-item representations, we argue that there may still exist avenues for further improvement. 
In this section, we first explain the difficulty of quantization in L$^2$Q-GCN$_{end}$, and then give an advanced solution with the annealing training strategy, namely L$^2$Q-GCN$_{anl}$, to enhance the recommendation performance. 

\subsection{Difficulty of Quantization in L$^2$Q-GCN$_{end}$}
To show it may be challenging to directly quantize L$^2$Q-GCN$_{end}$ for binary representation, we simulate the optimization trajectories of learnable embeddings and visually compare the loss landscapes of it with its full-precision version (excluding the quantization component) in Figure~\ref{fig:ste_loss}.
Concretely, following~\cite{nahshan2021loss, bai2020binarybert}, we manually assign perturbations to the learnable user-item embeddings as follows:
\begin{equation}
\boldsymbol{v}_u^{(l)} = \boldsymbol{v}_u^{(l)} \pm p \cdot \overline{|\boldsymbol{v}_u^{(l)}|} \cdot \boldsymbol{1}_u^{(l)}, \ \ \ \ \boldsymbol{v}_i^{(l)} = \boldsymbol{v}_i^{(l)} \pm p \cdot \overline{|\boldsymbol{v}_u^{(l)}|} \cdot \boldsymbol{1}_i^{(l)}, 
\end{equation}
where $\overline{|\boldsymbol{v}_u^{(l)}|}$ represents the absolute mean value of embedding $\boldsymbol{v}_u^{(l)}$ and perturbation magnitudes $p$ are from $\{0.01, 0.02, \cdots, 0.50\}$. $\boldsymbol{1}_u$ is an all-one vector. 
For each pair of perturbed user-item representations, we plot the loss distribution accordingly.

As we can observe, the full-precision version with no quantization produces a flat and smooth loss surface, showing the local convexity and thus easy to optimize.
On the contrary, L$^2$Q-GCN$_{end}$ has a bumping and complex loss landscape.
The steep loss curvature reflects L$^2$Q-GCN$_{end}$ is more sensitive to perturbation, showing the difficulty in quantization optimization.

\subsection{Upgrading in L$^2$Q-GCN$_{anl}$}
To alleviate the perturbation sensitivity and further improve the model performance, we propose L$^2$Q-GCN$_{anl}$. 

\subsubsection{\textbf{Quantization with Rescaling Approximation.}}
Our proposed L$^2$Q-GCN$_{anl}$ additionally includes layer-wise positive rescaling factors for each node, e.g., $\alpha_u^{(l)} \in \mathbb{R}^+$, such that $\boldsymbol{v}^{(l)}_u \approx$ $\alpha_u^{(l)} \boldsymbol{q}^{(l)}_u$.
In this work, we introduce a simple but effective approach to directly calculate these rescaling factors as: 
\begin{equation}
\alpha_u^{(l)} = \frac{||\boldsymbol{v}_u^{(l)}||_1}{d}, \ \ \alpha_i^{(l)} = \frac{||\boldsymbol{v}_i^{(l)}||_1}{d}.
\end{equation}
Instead of setting $\alpha_u^{(l)}$ as learnable, such deterministic computation substantially prunes the search space of parameters whilst attaining the approximation functionality. We demonstrate this in Section~\ref{sec:ablation}. 

Based on the quantized user-item representations and the corresponding rescaling factors, we have:
\begin{equation}
\resizebox{1\linewidth}{!}{$
\displaystyle
\mathcal{A}_u = \{\alpha_u^{(0)}\boldsymbol{q}^{(0)}_u, \alpha_u^{(1)}\boldsymbol{q}^{(1)}_u, \cdots, \alpha_u^{(L)}\boldsymbol{q}^{(L)}_u\}, \ \ \mathcal{A}_i = \{\alpha_i^{(0)}\boldsymbol{q}^{(0)}_i, \alpha_i^{(1)}\boldsymbol{q}^{(1)}_i, \cdots, \alpha_i^{(L)}\boldsymbol{q}^{(L)}_i\}. 
$}
\end{equation}
Consequently, L$^2$Q-GCN$_{anl}$ approximates $\mathcal{Q}_u$ and $\mathcal{Q}_i$ by $\mathcal{A}_u$ and $\mathcal{A}_i$, and updates Equations~\ref{eq:score} and \ref{eq:useQ} for model prediction accordingly.

\textbf{Space Cost Analysis.}
The total space cost of L$^2$Q-GCN$_{end}$ for storing the quantized user-item representations is $O((L+1)Nd)$ (or bits), where $N$ is the number of users and items and $d$ is the dimension of binary embeddings, e.g., $\boldsymbol{v}_u^{(l)}$. 
Furthermore, since L$^2$Q-GCN$_{anl}$ develops quantization with approximation, supposing that we use 32-bit floating-points for those rescaling factors, the space cost is $O((L+1)N(d+32))$ in total. 
Compared to the full-precision embedding table at each single one layer, e.g., 32-bit floating-point $\boldsymbol{v}_u^{{(l)}}$, $\mathcal{Q}_u$ and $\mathcal{Q}_i$ have the following theoretical compression ratios:
\begin{equation}
\label{eq:space}
\resizebox{1\linewidth}{!}{$
\displaystyle
ratio_{end} = \frac{32Nd}{(L+1)\cdot Nd} = \frac{32}{L+1}, \ \ ratio_{anl} = \frac{32Nd}{(L+1)\cdot N(d+32)} = \frac{32d}{(L+1)(d+32)}.
$}
\end{equation}
Normally, stacking too many layers will cause the \textit{over-smoothing} problem~\cite{li2018deeper,li2019deepgcns}, incurring performance detriment. Hence, A common setting for $L$ is $L<5$~\cite{lightgcn,ngcf,kipf2016semi,graphsage}, which can still achieve considerable embedding compression.

\subsubsection{\textbf{Annealing Training Strategy.}}
Another constructive design of L$^2$Q-GCN$_{anl}$ is the two-step \textit{annealing training strategy}: 
\begin{enumerate}[leftmargin=*]
\item we first mask the quantization function and train L$^2$Q-GCN$_{anl}$ with the full-precision embeddings;
\item when it converges to optimum, we trigger the quantization afterwards to find the targeted quantized representations.
\end{enumerate}

\begin{figure}[tp]
\begin{minipage}{0.5\textwidth}
\includegraphics[width=3.3in]{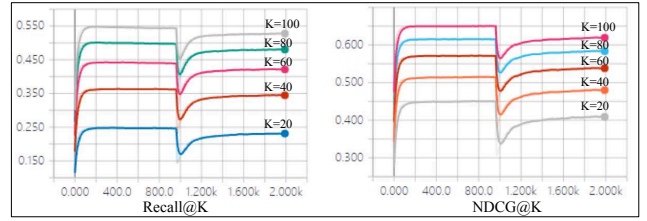}
\end{minipage} 
\setlength{\abovecaptionskip}{0.2cm}
\setlength{\belowcaptionskip}{0.2cm}
\caption{Annealing performance on MovieLens dataset.}
\label{fig:anneal}
\end{figure}

Intuitively, L$^2$Q-GCN$_{anl}$ moves from a tractable training space to the targeted one for quantization.
This can avoid unnecessary exploration towards different optimization directions at the beginning of quantization, guaranteeing the numerical stability in the whole model training.
Then at the middle period of training when triggering quantization, as presented in Figure~\ref{fig:anneal}, L$^2$Q-GCN$_{anl}$ firstly meets a performance retracement, but shortly afterwards, it recovers and continues to converge.
As we will demonstrate later in experiments, this straightforward but effective strategy can further produce better quantization representations with approximation for Top-K recommendation.

\textit{\underline{Clarification.}
we point out that the time cost of training L$^2$Q-GCN$_{anl}$ shares the same order of magnitude with L$^2$Q-GCN$_{end}$. 
In our implementation, we simply assign half of the total training epochs for the first step and leave the second half for quantization. 
One may also opt for more flexible strategy to trigger the quantization, e.g., early-stopping, or full-precision version pre-training.
}

So far, we have introduced all technical details of the proposed L$^2$Q-GCN$_{end}$ and L$^2$Q-GCN$_{anl}$.
For the corresponding pseudocodes, please refer to Appendix~\ref{sec:pseudocode}.
In the following section, we present the experimental results and analysis on our models.

\section{Experimental Results}
\label{sec:exp}
We evaluate our model on Top-K recommendation task with the aim of answering the following research questions:
\begin{itemize}[leftmargin=*]
\item \textbf{RQ1.} How does L$^2$Q-GCN perform compared to state-of-the-art full-precision and quantization-based recommender models?

\item \textbf{RQ2.} How is resource consumption of L$^2$Q-GCN? 

\item \textbf{RQ3.} How do proposed components of L$^2$Q-GCN$_{end}$ and L$^2$Q-GCN$_{anl}$ affect the performance?
\end{itemize}

\subsection{Dataset}
\begin{itemize}[leftmargin=*]
\item \textbf{MovieLens}\footnote{\url{https://grouplens.org/datasets/movielens/1m/}} is a widely adopted benchmark for movie recommendation. Similar to the setting in~\cite{hashgnn,he2016fast,chen2021modeling}, $y_{u,i} = 1$ if user $u$ has an explicit rating score towards item $i$, otherwise $y_{u,i} = 0$. In this paper, we use the MovieLens-1M data split.

\item \textbf{Gowalla}\footnote{\url{https://github.com/gusye1234/LightGCN-PyTorch/tree/master/data/gowalla}} is the check-in dataset~\cite{liang2016modeling} collected from Gowalla, where users share their locations by check-in. To guarantee the quality of the dataset, we extract users and items with no less than 10 interactions similar to~\cite{ngcf,hashgnn,lightgcn}. 

\item \textbf{Pinterest}\footnote{\url{https://sites.google.com/site/xueatalphabeta/dataset-1/pinterest_iccv}} is an implicit feedback dataset for image recommendation~\cite{geng2015learning}. Users and images are modeled in a graph. Edges represent the pins over images initiated by users. In this dataset, each user has at least 20 edges. We delete repeated edges between users and images to avoid data leakage in model evaluation.  

\item \textbf{Yelp2018}\footnote{\url{https://github.com/gusye1234/LightGCN-PyTorch/tree/master/data/yelp2018}} is collected from Yelp Challenge 2018 Edition. In this dataset, local businesses such as restaurants are treated as items. We retain users and items with over 10 interactions similar to~\cite{ngcf}.
\end{itemize}  

\begin{table}[h]
\setlength{\abovecaptionskip}{0.2cm}
\setlength{\belowcaptionskip}{0.2cm}
\centering
\caption{Datasets.}
\label{tab:datasets}
\setlength{\tabcolsep}{2.3mm}{
\begin{tabular}{|c | c  c  c  c|}
\toprule
           	 	&\# Users   & \# Items  & \# Interactions  &  Density\\
\midrule
\midrule
  MovieLens    	& 	{6,040} 		& {3,952}  		& 	{1,000,209}   	& 	{0.04190}\\
\midrule[0.1pt]
  Gowalla 		&	{29,858}		&	{40,981}	&	{1,027,370}		&	{0.00084}\\
\midrule[0.1pt]
  Pinterest 	&	{55,186}		&	{9,916}		&	{1,463,556}		&	{0.00267}\\
\midrule[0.1pt]
  Yelp2018 			&	{31,668}		&	{38,048}	&	{1,561,406}		&	{0.00130}\\
\bottomrule
\end{tabular}}
\end{table}

\subsection{Competing Methods}
We compare our model with two main streams of methods: (1) full-precision recommender systems including CF-based methods (NeurCF), and GCN-based models (NGCF, LightGCN), 
(2) 1-bit quantization-based models for general item retrieval tasks (LSH, HashNet) and for Top-K recommendation (HashGNN).

\begin{itemize}[leftmargin=*]
\item \textbf{LSH}~\cite{lsh} is a classical hashing method. LSH is firstly proposed to approximate the similarity search for massive high-dimensional data and we introduce it for Top-K recommendation by following the adaptation in~\cite{hashgnn}. 

\item \textbf{HashNet}~\cite{hashnet} is a state-of-the-art deep hashing method that is originally proposed for multimedia retrieval tasks.
Similar to~\cite{hashgnn}, we adapt it for graph data mainly by replacing the used AlexNet~\cite{krizhevsky2012imagenet} with the general graph convolutional network.

\item \textbf{HashGNN}~\cite{hashgnn} is the state-of-the-art 1-bit quantization-based recommender system method with GCN framework. We use HashGNN to denote the vanilla version with \textit{hard encoding} proposed in~\cite{hashgnn}, where each element of quantized user-item embeddings is strictly quantized. 
We use \textbf{HashGNN-soft} to represent the relaxed version proposed in~\cite{hashgnn}, where it adopts a Bernoulli random variable to provide the probability of replacing the quantized digits with continuous values in the original embeddings. 

\item \textbf{NeurCF}~\cite{neurcf} is one state-of-the-art neural network model for collaborative filtering. NeurCF models latent features of users and items to capture their nonlinear feature interactions.

\item \textbf{NGCF}~\cite{ngcf} is one of the state-of-the-art GCN-based recommender models. We compare L$^2Q$-GCN with NGCF mainly to study their performance capabilities in Top-K recommendation. 

\item \textbf{LightGCN}~\cite{lightgcn} is the latest state-of-the-art GCN-based recommendation model that has been widely evaluated. We include LightGCN in our experiment mainly to set up the benchmarking as a reference of full-precision recommendation capability.

\item \textbf{Quant-gumbel} is a variance of L$^2$Q-GCN with the implementation of Gumbel-softmax for quantization~\cite{gumbel1,gumbel2,zhang2019doc2hash}.
We first expand each embedding bit to a size-two one-hot encoding. Then Quant-gumbel utilizes the Gumbel-softmax trick to replace \textit{sign} function as relaxation for binary code generation.  
\end{itemize}

\begin{table*}[]
\setlength{\abovecaptionskip}{0.2cm}
\setlength{\belowcaptionskip}{0.2cm}
\centering
  \caption{Performance comparison (underline represents the best performing model; R and D refer to Recall and NDCG). }
  \label{tab:top20}
  \setlength{\tabcolsep}{0.6mm}{
  \begin{tabular}{|c|c c c c|c c c c|c c c c|c c c c|} 
    \toprule
    \multirow{2}*{Model} & \multicolumn{4}{c|}{MovieLens (\%)} & \multicolumn{4}{c|}{Gowalla (\%)} & \multicolumn{4}{c|}{Pinterest (\%)} & \multicolumn{4}{c|}{Yelp2018 (\%)} \\
        ~ & R@20 & R@100 & N@20 & N@100 & R@20 & R@100 & N@20 & N@100 & R@20 & R@100 & N@20 & N@100 & R@20 & R@100 & N@20 & N@100\\
    \midrule
    \midrule
    LSH             & {9.35}  & {23.41}  & {13.45}   & {33.49}   & {6.23}  & {15.69}   & {10.23}  & {19.23}  & {6.41}  & {16.55}  & {8.56}   & {15.32}   & {2.73}  & {8.77}   & {4.73}  & {11.63} \\
    HashNet         & {14.98}  & {35.67}  & {23.41}   & {38.14}   & {10.11}  & {24.19}   & {15.31}  & {23.11}  & {8.93}  & {29.43}  & {10.37}   & {21.63}   & {2.64}  & {9.45}   & {6.42}  & {14.68} \\
    HashGNN         & {12.55}  & {34.54}  & {23.63}   & {36.54}   & {9.63}  & {22.13}   & {15.85}  & {24.01}  & {8.01}  & {25.27}  & {10.48}   & {21.19}   & {3.22}  & {10.96}   & {6.62}  & {14.61} \\
    HashGNN-soft    & {18.54}  & {41.83}  & {36.56}   & {55.16}   & {11.49}  & {25.88}   & {17.84}  & {26.53}  & {10.87}  & {34.14}  & {12.35}   & {24.99}   & {4.29}  & {14.03}   & {8.30}  & {17.73} \\
    Quant-gumbel     & {17.48} & {42.08} & {34.57} & {56.50} & {10.78} & {26.58} & {15.62} & {25.08} & {9.78} & {30.63} & {11.15} & {22.92} & {3.91} & {13.00} & {8.07} & {17.08}\\
    \midrule[0.1pt]
    NeurCF           & {19.43}  & {48.04}  & {36.85}   & {51.71}   & {13.95}  & {32.76}   & {22.21}  & {27.39}  & {9.09}  & {29.52}  & {13.55}   & {22.24}   & {3.12}  & {10.97}   & {6.45}  & {14.91} \\
    NGCF             & {24.35}  & {54.92}  & {37.30}   & {53.64}   & {15.53}  & {33.32}   & {23.21}  & {28.44}  & {14.30}  & {39.74}  & {13.01}   & {27.06}   & {5.45}  & {15.32}   & {8.73}  & {19.33} \\

    {LightGCN}        & {\underline{25.01}}  & {\underline{55.71}}  & {\underline{44.73}}   & {\underline{65.09}}   & {\underline{17.80}}  & {\underline{37.02}}   & {\underline{24.76}}  & {\underline{34.91}}  & {\underline{14.78}}  & {\underline{39.98}}  & {\underline{15.91}}   & {\underline{28.93}}   & {\underline{6.11}}  & {\underline{18.05}}   & {\underline{10.95}}  & {\underline{21.59}} \\
    \midrule
    \midrule 
    \textbf{L$^2$Q-GCN$_{end}$} & {20.52}  & {49.27}  & {38.41}   & {60.03}   & {14.62}  & {32.24}   & {21.24}  & {30.97}  & {12.52}  & {35.67}  & {13.92}   & {26.37}   & {5.10}  & {16.31}   & {9.62}  & {19.88} \\
    \% Capacity     & \textit{\small 82.05\%}  & \textit{\small88.44\%}  & \textit{\small85.87\%}   & \textit{\small92.23\%}   & \textit{\small82.13\%}  & \textit{\small87.10\%}   & \textit{\small85.78\%}  & \textit{\small88.71\%}  & \textit{\small84.71\%}  & \textit{\small89.22\%}  & \textit{\small87.49\%}   & \textit{\small91.15\%}   & \textit{\small83.47\%}  & \textit{\small90.36\%}   & \textit{\small87.85\%}  & \textit{\small92.08\%} \\
    \textbf{L$^2$Q-GCN$_{anl}$} & {22.81}  & {51.96}  & {42.44}   & {62.96}   & {16.12}  & {34.39}   & {23.62}  & {33.52}  & {13.87}  & {38.25}  & {15.31}   & {28.13}   & {5.74}  & {17.63}   & {10.67}  & {21.32} \\
    \% Capacity     & \textit{\small91.20\%}  & \textit{\small93.27\%}  & \textit{\small94.88\%}   & \textit{\small96.73\%}   & \textit{\small90.56\%}  & \textit{\small92.90\%}   & \textit{\small95.40\%}  & \textit{\small96.02\%}  & \textit{\small93.84\%}  & \textit{\small95.67\%}  & \textit{\small96.23\%}   & \textit{\small97.23\%}   & \textit{\small93.94\%}  & \textit{\small97.67\%}   & \textit{\small97.44\%}  & \textit{\small98.75\%} \\
    \bottomrule
  \end{tabular}}
\end{table*}

\subsection{Experiment Setup}
\label{sec:exp_setup}
In the evaluation of Top-K recommendation, we apply the learned user-item representations to rank $K$ items for each user with the highest predicted scores, i.e., $\hat{y}_{u,i}$. We choose two widely-used evaluation protocols Recall@$K$ and NDCG@$K$ to evaluate Top-K recommendation capability. 

We implement L$^2$Q-GCN model under Python 3.7 and PyTorch 1.14.0 with non-distributed training. 
The experiments are run on a Linux machine with 4 NVIDIA V100 GPU, 4 Intel Core i7-8700 CPUs, 32 GB of RAM with 3.20GHz.
For all the baselines, we follow the official hyper-parameter settings from original papers or as default in corresponding codes. 
For methods lacking recommended settings, we apply a grid search for hyper-parameters.
The embedding dimension is searched in \{$32, 64, 128, 256, 512$\}. 
The learning rate $\eta$ is tuned within \{$10^{-3}, 5\times10^{-3}, 10^{-2}, 5\times10^{-2}$\} and the coefficient of $L2$ normalization $\lambda$ is tuned among \{$10^{-5}, 10^{-4}, 10^{-3}$\}. 
We initialize and optimize all models with default normal initializer and Adam optimizer~\cite{adam}. 
To guarantee reproducibility, we report all the hyper-parameter settings in Appendix~\ref{sec:parameter}.

\subsection{Performance Analysis (RQ1)}

In this section, we present a comprehensive performance analysis between L$^2Q$-GCN with two layers and competing recommender models of full-precision-based and quantization-based.
We evaluate Top-K recommendation over four datasets by varying K in \{20, 40, 60, 80, 100\}. 
To achieve a more detailed performance comparison, we summarize the results of Top@20 and Top@100 recommendation in Table~\ref{tab:top20}. 
We also curve their complete results of Recall@K and NDCG@K metrics and attach them in Appendix~\ref{sec:topk}. 
Generally, L$^2Q$-GCN has made great improvements over quantization-based models and shows the competitive performance compared to full-precision models.
We have the following observations:
\begin{itemize}[leftmargin=*]
\item \textbf{The results demonstrate the superiority of L$^2Q$-GCN over all quantization-based models.}
(1) As shown in Table~\ref{tab:top20}, the state-of-the-art quantization-based GCN model, i.e., HashGNN (and HashGNN-soft), works better than traditional quantization-based baselines, e.g., LSH, HashNet. 
This shows the effectiveness of \textit{graph convolutional} architecture in capturing latent information within interaction graphs for quantization preparation and indicates that a direct adaptation of conventional quantization methods may not well handle the Top-K recommendation task.\\
(2) Furthermore, thanks to our proposed quantization-based graph convolution design, both L$^2Q$-GCN$_{end}$ and L$^2Q$-GCN$_{anl}$ consistently outperform HashGNN and its relaxed version HashGNN-soft.
The main reason is that, our topology-aware quantization significantly enriches the user-item representations and alleviates the feature smoothing issue caused by the numerical quantization.
We conduct the ablation study on this in the later section.

\item \textbf{Compared to full-precision models, L$^2Q$-GCN presents a competitive performance recovery.}
(1) L$^2Q$-GCN$_{end}$ shows the performance superiority over traditional collaborative filtering model, i.e., NeurCF.
Considering the large improvement of two GCN-based models, i.e., NGCF and LightGCN, against NeurCF, we can infer the performance gap between L$^2Q$-GCN$_{end}$ and NeurCF mainly comes from the proposed quantization-based graph convolution. 
(2) Compared to the best model LightGCN, taking Recall metric as an example,  L$^2Q$-GCN$_{end}$ shows about $82$$\sim$$85\%$ and $87$$\sim$$90\%$ performance capacity in terms of Top@20 and Top@100.
This indicates that, as the value of K increases, L$^2Q$-GCN$_{end}$ can further improve the recommendation accuracy and narrow the gap to LightGCN.
(3) Moreover, by utilizing the quantization approximation and annealing training strategy, L$^2Q$-GCN$_{anl}$ can achieve better performance than NGCF on Gowalla and Yelp2018 datasets.
Compared to our basic implementation L$^2Q$-GCN$_{end}$, L$^2Q$-GCN$_{anl}$ further improves the performance recovery by $8$$\sim$$10\%$ and $9$$\sim$$10\%$ in terms of Recall@20 and Recall@100 across all benchmarks, proving the effectiveness of our proposed modification in L$^2Q$-GCN$_{anl}$.
(4) In addition, with K increasing up to 100, L$^2Q$-GCN$_{anl}$ presents a similar trend with L$^2Q$-GCN$_{end}$ such that it performs even closely to LightGCN, i.e., $93$$\sim$$98\%$ and $96$$\sim$$99\%$ in terms of Recall and NDCG, respectively.
In a nutshell, the prediction capability of both L$^2Q$-GCN$_{end}$ and L$^2Q$-GCN$_{anl}$ develops from fine-grained ranking tasks to coarse-grained ones, e.g., Top-20 to Top-100 recommendation.

\item \textbf{Both L$^2Q$-GCN$_{end}$ and L$^2Q$-GCN$_{anl}$ show the deployment flexibility towards different application scenarios.}
In the pipeline of industrial recommender systems, \textit{recall} and \textit{re-ranking} are two important stages that substantially influence recommendation quality. 
\textit{Recall} refers to the process of quickly retrieving candidate items from the whole item pool that a given user may interest.
Based on the more complex scoring algorithms, \textit{re-ranking} outputs a precise ranking list of candidate items. 

On the one hand, as we will show later, L$^2Q$-GCN$_{end}$ can speed up about 40\% for candidate generation. Considering its performance improvement on coarse-grained ranking tasks, e.g., Top-100 recommendation, L$^2Q$-GCN$_{end}$ actually provides an alternative option to accelerate the \textit{recall} stage.
On the other hand, L$^2Q$-GCN$_{anl}$ further optimizes the recommendation accuracy that performs similarly to LightGCN.
Since L$^2Q$-GCN$_{anl}$ reduces the embedding storage cost to about 9$\times$, we can treat L$^2Q$-GCN$_{anl}$ as a substitute for the best model, i.e., LightGCN, as a trade-off between space cost and prediction accuracy.
We report the details of space and time cost for embedding storage and online inference in the following section.
\end{itemize}

\subsection{Resource Consumption Analysis (RQ2)}
\begin{figure}[t]
\begin{minipage}{0.5\textwidth}
\includegraphics[width=3.2in]{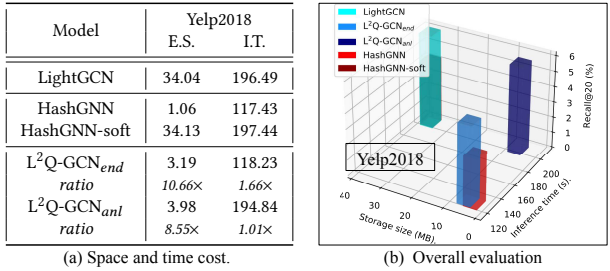}
\end{minipage} 
\setlength{\abovecaptionskip}{0.2cm}
\setlength{\belowcaptionskip}{0.2cm}
\caption{Results of two-layer networks with 128-dimension embeddings (E.S. and I.T. are the abbreviations of embedding size (MB) and inference time (s). Best view in color).}
\label{fig:tradeoff}
\end{figure}

In this section, we study the resource consumption in embedding storage and inference time cost. 
We compare our models with the state-all-the-art full-precision model and quantization-based model, i.e., LightGCN and HashGNN.
We take their two-layer structures with the same 128-dimensional embeddings as reference and illustrate with the largest dataset, i.e., Yelp2018, in Figure~\ref{fig:tradeoff}(a). 
Observations in this section can be popularized to other three datasets and we report the complete results in Appendix~\ref{sec:topk}. 

\textbf{Embedding storage compression.}
Quantized embeddings can largely reduce the space consumption for offline disk storage.
To measure the embedding size, we save these embeddings to the disk such that they can recover the well-trained user-item representations for inference.
As we can observe, after the binarization for user-item embeddings, L$^2$Q-GCN$_{end}$ and L$^2$Q-GCN$_{anl}$ can achieve the space reduction with a factor of about 11$\times$ and 9$\times$, respectively. 
This basically follows the theoretical bounds that are computed in Equation~\ref{eq:space}, i.e., $ratio_{end}$ and $ratio_{anl}$ when $L=2$.
Furthermore, considering the performance improvement, the space usage of approximation factors of L$^2$Q-GCN$_{anl}$ is actually acceptable, which may dispel the concerns of large additional storage overhead.

\textbf{Online inference acceleration.}
We evaluate the time cost including the score estimation and sorting.
L$^2$Q-GCN predicts the scores between users and items by conducting embedding multiplications.
At the stage of online inference, we first interpret these binary embeddings by signed integers, e.g., int8, and then conduct integer arithmetics including integer summations and matrix multiplications.
Please notice that we leave the development of bitwise operations for online inference for future work.
To give a fair comparison on the inference time cost, we disable all arithmetic optimization such as BLAS, MKL, and conduct the experiments using the vanilla NumPy provided by~\footnote{\url{https://www.lfd.uci.edu/~gohlke/pythonlibs/}}.
As shown in Figure~\ref{fig:tradeoff}(a), purely based on the quantized embeddings, L$^2$Q-GCN$_{end}$ can achieve 39.83\% (118.23$\rightarrow$196.49) of inference acceleration. 
L$^2$Q-GCN$_{anl}$ takes a similar running time with LightGCN as it introduces the approximation factors in score estimation. 
Furthermore, Figure~\ref{fig:tradeoff}(b) visualizes the overall evaluation in terms of resource consumption and recommendation accuracy.
As the cube's front-high corner means the ideal optimal performance, our proposed methods make a good balance w.r.t consumption and accuracy.


\subsection{Ablation Study of L$^2$Q-GCN Model (RQ3)}
\label{sec:ablation}
We evaluate the necessity of each model component in both L$^2$Q-GCN$_{end}$ and L$^2$Q-GCN$_{anl}$. Due to the page limits, we report the Top-20 recommendation results as a reference in Table~\ref{tab:ablation}.

\subsubsection{\textbf{Effect of Topology-aware Quantization.}}
To substantiate the impact of topology-aware quantization in graph convolution, we give a variant of L$^2$Q-GCN$_{end}$, i.e., \textsl{w/o TQ}, by disabling the layer-wise quantization and setting it as the final encoder of full-precision graph convolution.
As we can observe in Table~\ref{tab:ablation}, variant \textsl{w/o TQ} remarkably underperforms L$^2$Q-GCN$_{end}$.
This demonstrates that simply using the latest-updated embeddings from the GCN framework may not sufficiently model the unique latent features of both users and items, especially for the quantization-based ranking. 
Via capturing the intermediate information for representation enrichment, our topology-aware quantization can effectively alleviate the ranking smoothness issue caused by the limited expressivity of discrete embeddings.
This helps to make the quantized representations of both users and items more discriminative, which leads to the performance improvement on Top-K recommendation.

\subsubsection{\textbf{Effect of Multi-loss in Optimization.}}
To study the effect of BPR loss $\mathcal{L}_{bpr}$ and graph reconstruction loss $\mathcal{L}_{rec}$, we set two variants, termed by \textsl{w/o $\mathcal{L}_{bpr}$} and \textsl{w/o $\mathcal{L}_{rec}$}, to optimize L$^2$Q-GCN$_{end}$ separately.
As shown in Table~\ref{tab:ablation}, with all other model components, partially using one of $\mathcal{L}_{bpr}$ and $\mathcal{L}_{rec}$ produces large performance decay to L$^2$Q-GCN$_{end}$. 
This confirms the effectiveness of our proposed muti-loss design:
while $\mathcal{L}_{bpr}$ assigns higher prediction values to observed interactions, 1.e., $y_{u,i}=1$, than the unobserved user-item pairs, 
$\mathcal{L}_{rec}$ transfers the graph reconstruction problem to a classification task by using the full-precision embeddings in training.
By collectively optimizing these two loss functions, L$^2$Q-GCN$_{end}$ can learn precise intermediate embeddings from $\mathcal{L}_{rec}$, and produce quantized representations with high-quality relative order information regularized by $\mathcal{L}_{bpr}$ accordingly.

\subsubsection{\textbf{Effect of Rescaling Approximation.}}
We now discuss the effect of approximation factors in L$^2$Q-GCN$_{anl}$.
We create two variants, namely \textsl{w/o RAF} and \textsl{w/in LF}.
\textsl{w/o RAF} directly removes the rescaling approximation factors and \textsl{w/in LF} means replacing our original approximation factors with learnable ones. 
We optimize both variants \textsl{w/o RAF} and \textsl{w/in LF} with the annealing training strategy.
(1) The performance decline of \textsl{w/o RAF} proves the effectiveness of rescaling approximation for user-item representations.
Although these factors are directly calculated and may not be theoretically optimal, they reflect the numerical uniqueness of embeddings for both users and items, which substantially improves L$^2$Q-GCN$_{anl}$'s prediction capability. 
(2) As for \textsl{w/in LF}, the design of learnable rescaling factors does not achieve good performance as expected. 
One explanation is that, our proposed model currently does not post a direct mathematical constraint to learnable factors ($lf$), e.g., $lf_u^{(l)} = \argmin(\boldsymbol{v}^{(l)}_u$, $lf_u^{(l)} \boldsymbol{q}^{(l)}_u)$, mainly because they have different embedding dimensionality.
This means that purely relying on the stochastic optimization may hardly reach the optimum.
In a word, considering the additional search space introduced by this regularization term, we argue that our deterministic rescaling method is simple but effective in practice.

\begin{table}[t]
\setlength{\abovecaptionskip}{0.2cm}
\setlength{\belowcaptionskip}{0.2cm}
\centering
\caption{Ablation study.}
\label{tab:ablation}
\setlength{\tabcolsep}{0.3mm}{
\begin{tabular}{|c |c c|c c|c c|c c|}
\toprule
 \multirow{2}*{Variant} & \multicolumn{2}{c|}{MovieLens} & \multicolumn{2}{c|}{Gowalla} & \multicolumn{2}{c|}{Pinterest} & \multicolumn{2}{c|}{Yelp2018} \\
               ~  & R@20 & N@20 & R@20 & N@20 & R@20 & N@20 & R@20 & N@20\\
\midrule
\midrule
\multicolumn{9}{|c|}{L$^2$Q-GCN$_{enl}$} \\ 
\midrule
  \multirow{2}*{\textsl{w/o TQ}}    &{17.73}& {35.31}   & {11.63}& {15.58}  & {10.29}& {11.60}  & {4.33} & {8.58} \\
  ~        &\textit{\small{-13.60\%}}  &\textit{\small{-8.07\%}}  &\textit{\small{-20.45\%}}  &\textit{\small{-26.65\%}}  &\textit{\small{-17.81\%}}  &\textit{\small{-16.67\%}}  &\textit{\small{-15.10\%}}  &\textit{\small{-10.81\%}} \\
  \midrule[0.1pt]
  \multirow{2}*{\textsl{w/o $\mathcal{L}_{bpr}$}}    &{19.67}& {37.22}   & {8.66}& {13.33}  & {5.12}& {6.01}  & {3.52}& {7.33} \\
  ~        &\textit{\small{-4.14\%}}  &\textit{\small{-3.10\%}}  &\textit{\small{-40.77\%}}  &\textit{\small{-37.24\%}}  &\textit{\small{-59.11\%}}  &\textit{\small{-56.82\%}}  &\textit{\small{-30.98\%}}  &\textit{\small{-23.80\%}} \\
 \midrule[0.1pt]
  \multirow{2}*{\textsl{w/o $\mathcal{L}_{rec}$}}   &{16.98}& {32.76}   & {8.32}& {9.28}  & {10.86}& {11.55}  & {3.33}& {6.60} \\
  ~        &\textit{\small{-17.25\%}}  &\textit{\small{-14.71\%}}  &\textit{\small{-43.09\%}}  &\textit{\small{-56.31\%}}  &\textit{\small{-13.26\%}}  &\textit{\small{-17.03\%}}  &\textit{\small{-34.71\%}}  &\textit{\small{-31.39\%}} \\
  \midrule[0.1pt]
  \textbf{Best}   &\textbf{20.52}& \textbf{38.41}   & \textbf{14.62}& \textbf{21.24}  & \textbf{12.52}& \textbf{13.92}  & \textbf{5.10}& \textbf{9.62} \\
\midrule
\multicolumn{9}{|c|}{L$^2$Q-GCN$_{anl}$} \\ 
\midrule
\multirow{2}*{\textsl{w/o RAF}}    &{20.86}& {38.14}   & {10.29}& {12.10}  & {11.19}& {12.11}  & {4.25}& {8.09} \\
~        &\textit{\small{-8.55\%}}  &\textit{\small{-10.13\%}}  &\textit{\small{-36.17\%}}  &\textit{\small{-48.77\%}}  &\textit{\small{-19.32\%}}  &\textit{\small{-20.90\%}}  &\textit{\small{-25.96\%}}  &\textit{\small{-24.18\%}} \\
\midrule[0.1pt]
\multirow{2}*{\textsl{w/in LF}}    &{20.05}& {38.25}   & {14.53}& {21.23}  & {12.35}& {13.65}  & {5.56}& {10.20} \\
~        &\textit{\small{-12.10\%}}  &\textit{\small{-9.87\%}}  &\textit{\small{-9.86\%}}  &\textit{\small{-10.12\%}}  &\textit{\small{-10.96\%}}  &\textit{\small{-10.84\%}}  &\textit{\small{-3.13\%}}  &\textit{\small{-4.40\%}} \\
\midrule[0.1pt]
\multirow{2}*{\textsl{w/o AT}}    &{21.24}& {40.05}   & {15.09}& {22.70}  & {13.37}& {14.75}  & {5.27}& {9.96} \\
  ~        &\textit{\small{-6.88\%}}  &\textit{\small{-5.63\%}}  &\textit{\small{-6.39\%}}  &\textit{\small{-3.90\%}}  &\textit{\small{-3.60\%}}  &\textit{\small{-3.66\%}}  &\textit{\small{-8.19\%}}  &\textit{\small{-6.65\%}} \\
  \midrule[0.1pt]
\textbf{Best}   &\textbf{22.81}& \textbf{42.44}   & \textbf{16.12}& \textbf{23.62}  & \textbf{13.87}& \textbf{15.31}  & \textbf{5.74}& \textbf{10.67} \\
\bottomrule
\end{tabular}}
\end{table}

\subsubsection{\textbf{Effect of Annealing Training Strategy.}}
We disable the annealing training strategy by only adapting the rescaling approximation design to L$^2$Q-GCN$_{end}$ and denote the variant as \textsl{w/o AT}.
The performance of \textsl{w/o AT} well demonstrate the usefulness of our utilized annealing training strategy in avoiding unnecessary optimization directions and the numerical stability for producing better recommendation accuracy. 



In conclusion, the ablation study well justifies the necessity of each model component in both L$^2$Q-GCN$_{end}$ and L$^2$Q-GCN$_{anl}$.
We also discuss the effect of different hyperparameter settings, e.g., layer depth $L$, quantization dimension $d$, to model performance and attached the results in Appendix~\ref{sec:topk}.

\section{Conclusion and Future Work}
\label{sec:con}
In this work, we propose L$^2$Q-GCN to study the problem of 1-bit representation quantization for Top-K recommendation.
While L$^2$Q-GCN$_{end}$ implements the basic framework of L$^2$Q-GCN to generate the quantized user-item representations, L$^2$Q-GCN$_{anl}$ further improves the recommendation capability by attaining 90$\sim$99\% performance recovery compared to the state-of-the-art model.
The extensive experiments over four real benchmarks not only prove the effectiveness of our proposed models but also justify the necessity of each model component.

As for future work, we point out two possible directions. (1) Instead of relying on integer arithmetics, how to develop the bitwise-operation-supported computation for efficient inference is an important topic to investigate. 
(2) It is also worth studying the problem of complete network binarization for the GCN framework, as it is more fundamental to many GCN-related methods for model compression and computation acceleration.

\bibliographystyle{ACM-Reference-Format}
\balance
\bibliography{ref}

\balance
\clearpage
\appendix
\setcounter{table}{0}
\setcounter{figure}{0}

\section{Notation and Meanings}
\label{sec:notation}
Table~\ref{tab:notation} explains all key notations used in this paper.

\section{pseudocodes of L$^2$Q-GCN}
\label{sec:pseudocode}
We attached the detailed pseudocodes of L$^2$Q-GCN$_{end}$ and L$^2$Q-GCN$_{anl}$ in Algorithms~\ref{alg:end} and \ref{alg:anl}, respectively.

\section{Hyper-parameter Settings}
\label{sec:parameter}
We report all the hyper-parameter settings in Table~\ref{tab:hyperparameter}.
Please notice that, in table~\ref{tab:hyperparameter}, $B$ is the batch size, $\eta$ represents the learning rate, and $\lambda$ is the coefficient of $L$2 normalization.

\section{Complete Results}
\label{sec:topk}
\begin{figure*}[t]
\begin{minipage}{1\textwidth}
\includegraphics[width=7in]{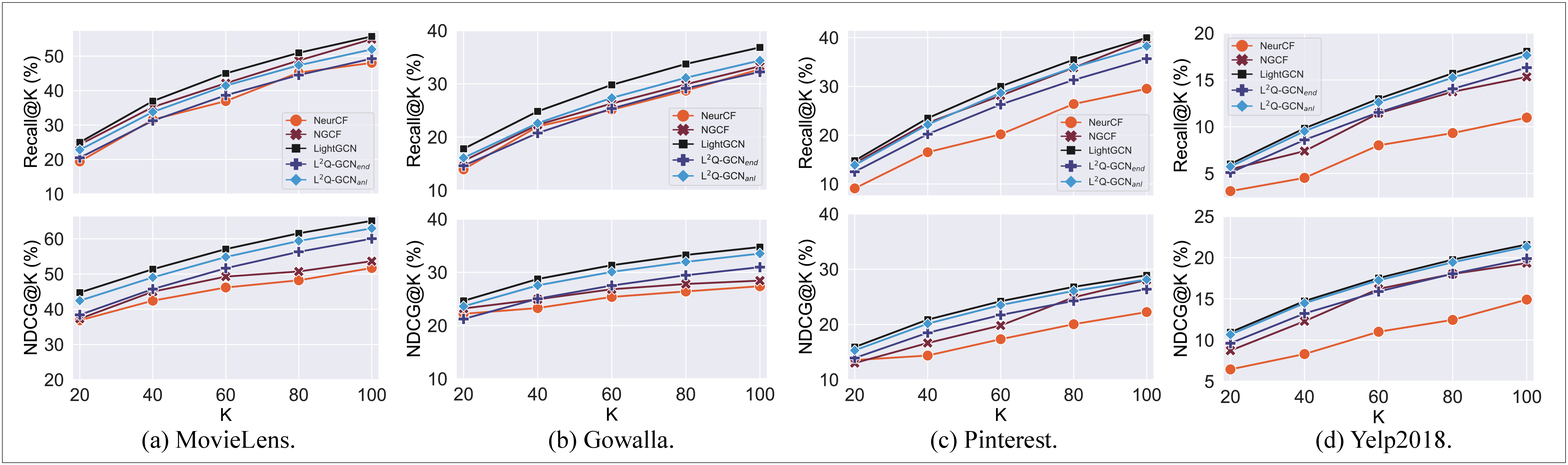}
\end{minipage} 
\caption{topk.}
\label{fig:topk}
\end{figure*}

For detailed curves of Top-K recommendation, please refer to Figure~\ref{fig:topk}.
For the study of different hyperparameter setttings, i.e., layer depth $L$ and quantization dimension $d$, please refer to Table~\ref{tab:layer} and Table~\ref{tab:dimension} accordingly.

\begin{table}[bp]
\setlength{\abovecaptionskip}{0cm}
\setlength{\belowcaptionskip}{0cm}
\caption {Notations and meanings. }
\label{tab:notation}
  \footnotesize
  \begin{tabular}{|c|l|} 
     \hline
          {\bf Notation} & {\bf Meaning}\\
     \hline\hline
          $\mathcal{U}$, $\mathcal{I}$  & \tabincell{c}{The sets of users, items.}\\
    \hline
         \tabincell{l}{$y_{u,i}$,\\ $\hat{y}_{u,i}$}      & \tabincell{l}{$y_{u,i}$$=$$1$ means there is an observed interaction between user $u$ and \\ 
         item $i$, otherwise $y_{u,i}$$=$$0$. $\hat{y}_{u,i}$ is the estimated matching score. } \\
    \hline
          $\boldsymbol{v}_x^{(l)}$,  & continuous embedding of node $x$ at the $l$-th layer.\\
    \hline
          $\boldsymbol{q}_x^{(l)}$,  & binary embedding of node $x$ at the $l$-th layer.\\
    \hline
          $\mathcal{Q}_x$,  & quantized representation of node $x$.\\
    \hline
          $\mathcal{A}_x$,  & approximation-based quantized representation of node $x$.\\
    \hline
  \end{tabular}
\end{table}

\begin{algorithm}[bp]
\small
\caption{L$^2$Q-GCN$_{end}$ algorithm}
\label{alg:end}
\LinesNumbered  
\KwIn{Interaction graph $\mathcal{G}$; trainable parameters {\footnotesize $\Theta$: $\{\boldsymbol{v}_{u}\}_{u\in\mathcal{U}}$, $\{\boldsymbol{v}_i\}_{i\in\mathcal{I}}$, $\boldsymbol{W}$}; hyper-parameters: {\footnotesize $B$, $d_{end}$, $L$, $\eta$, $\lambda$. } }

\KwOut{Prediction function $\mathcal{F}(u,i|\Theta, \mathcal{G})$} 
$\mathcal{Q}_u \gets \emptyset$, $\mathcal{Q}_i \gets \emptyset$;\\
\While{\rm{L$^2$Q-GCN$_{end}$ not converge}}{
    \For{$(u,i) \in \mathcal{G}$ \rm{that} $y_{u,i}=1$}{
        \For{$l = 1, \cdots, L$}{
          $\boldsymbol{v}_{u}^{(l)} \gets \sum_{i\in \mathcal{N}(u)} \frac{1}{\sqrt{|\mathcal{N}(u)|\cdot|\mathcal{N}(i)|}}v^{(l-1)}_i$; \\
          $\boldsymbol{v}_{i}^{(l)} \gets \sum_{u\in \mathcal{N}(i)} \frac{1}{\sqrt{|\mathcal{N}(i)|\cdot|\mathcal{N}(u)|}}v^{(l-1)}_u$; \\
          $q_u^{(l+1)} \gets \sign\big(\boldsymbol{W}^T\boldsymbol{v}^{(l)}_u\big)$; $q_i^{(l)} \gets \sign\big(\boldsymbol{W}^T\boldsymbol{v}^{(l)}_i\big)$; \\
          Update ($\mathcal{Q}_u$, $\mathcal{Q}_i$);\\
          $\frac{\partial \boldsymbol{q}^{(l)}_u}{\partial \phi}, \frac{\partial \boldsymbol{q}^{(l)}_i}{\partial \phi} \gets \boldsymbol{1}_{|\phi| \leq 1}$ for backforward propagation; \\
        }
      Update ($\mathcal{Q}_u$, $\mathcal{Q}_i$) with $q_u^{(0)}, q_i^{(0)}$;\\
      $\mathcal{L} \gets $ compute loss and optimize L$^2$Q-GCN$_{end}$ model;\\ 
    }
}
\KwRet $\mathcal{F}$.\\
\end{algorithm}

\begin{table}[bp]
\setlength{\abovecaptionskip}{0cm}
\setlength{\belowcaptionskip}{0cm}
\centering
\caption{Hyper-parameter settings for the four datasets.}
\label{tab:hyperparameter}
\setlength{\tabcolsep}{2.8mm}{
\begin{tabular}{|c | c  c  c  c|}
\toprule
                  & MovieLens   & Gowalla    & Pinterest     & Yelp2018  \\
\midrule 
\midrule
  $B$             &  2048       &  	2048     & 2048        &  2048   \\
  $d_{end}$		    &  256		    &	  64		   &  64		 & 64 		\\
  $d_{anl}$       &  256        &   256      &  128             & 256     \\
  $\eta$          & $1\times10^{-3}$      & $1\times10^{-3}$      & $1\times10^{-3}$        & $1\times10^{-3}$      \\
  $\lambda$       & $1\times10^{-4}$      & $1\times10^{-4}$      & $1\times10^{-4}$        & $1\times10^{-4}$      \\
\bottomrule
\end{tabular}}
\end{table}

\begin{figure*}[t]
\begin{minipage}{1\textwidth}
\includegraphics[width=7in]{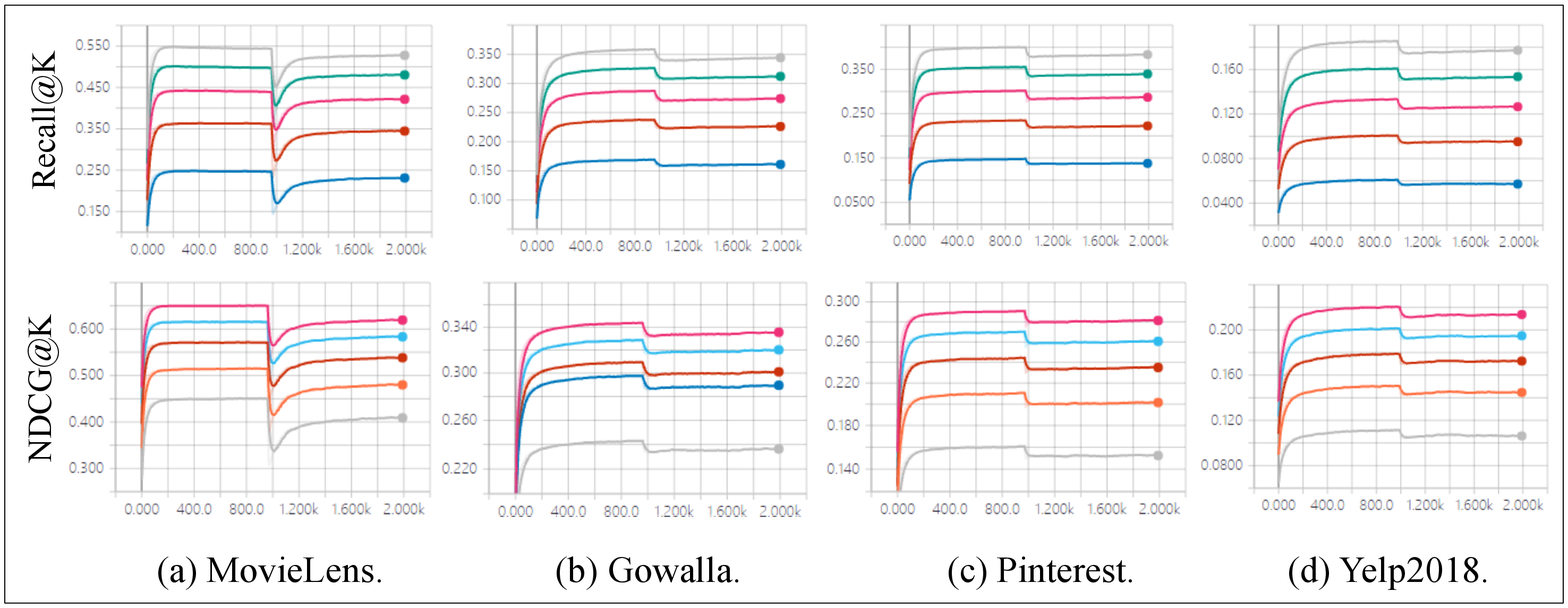}
\end{minipage} 
\caption{Complete performance curves with annealing tranining strategy.}
\label{fig:complete_annealing}
\end{figure*}

\begin{table}[bp]
\setlength{\abovecaptionskip}{0cm}
\setlength{\belowcaptionskip}{0cm}
\centering
\caption{Space and time results of two-layer networks with 128-dimension embeddings.}
\label{tab:efficiency}
\setlength{\tabcolsep}{0.3mm}{
\begin{tabular}{|c |c c|c c|c c|c c|}
\toprule
 \multirow{2}*{Model} & \multicolumn{2}{c|}{MovieLens} & \multicolumn{2}{c|}{Gowalla} & \multicolumn{2}{c|}{Pinterest} & \multicolumn{2}{c|}{Yelp2018} \\
               ~  & E.S. & I.T. & E.S. & I.T. & E.S. & I.T. & E.S. & I.T.\\
\midrule
\midrule
  LightGCN        &{4.88}& {3.42}   & {34.59}& {190.41}   & {28.45}& {82.66}   & {34.04}& {196.49} \\
\midrule
  L$^2$Q-GCN$_{end}$    &{0.47}& {2.19}   & {3.24}& {118.97}  & {2.98}& {53.48}  & {3.19}& {118.23} \\
  \textit{Gap}         &\textit{\small{10.38$\times$}} &\textit{\small{1.56$\times$}} &\textit{\small{10.68$\times$}} &\textit{\small{1.60$\times$}} &\textit{\small{9.55$\times$}} &\textit{\small{1.55$\times$}} &\textit{\small{10.66$\times$}} &\textit{\small{1.66$\times$}} \\
  L$^2$Q-GCN$_{anl}$    &{0.58}& {3.55}   & {4.05}& {191.32}  & {3.73}& {81.77}  & {3.98}& {194.84} \\
  \textit{Gap}         &\textit{\small{8.41$\times$}} &\textit{\small{0.96$\times$}} &\textit{\small{8.54$\times$}} &\textit{\small{1.00$\times$}} &\textit{\small{7.63$\times$}} &\textit{\small{1.01$\times$}} &\textit{\small{8.55$\times$}} &\textit{\small{1.01$\times$}} \\
\bottomrule
\end{tabular}}
\end{table}

\begin{table}[bp]
\setlength{\abovecaptionskip}{0cm}
\setlength{\belowcaptionskip}{0cm}
\centering
\caption{Top-20 recommendation (\%) w.r.t different $L$.}
\label{tab:layer}
\setlength{\tabcolsep}{0.7mm}{
\begin{tabular}{|c | c c | c c | c c | c c |}
\toprule
\multirow{2}*{$L$}  & \multicolumn{2}{c|}{MovieLens} & \multicolumn{2}{c|}{Gowalla} & \multicolumn{2}{c|}{Pinterest} & \multicolumn{2}{c|}{Yelp2018} \\
               ~  & R@20 & N@20 & R@20 & N@20 & R@20 & N@20 & R@20 & N@20\\
\midrule
\midrule
                \multicolumn{9}{|c|}{L$^2$Q-GCN$_{enl}$} \\
\midrule[0.1pt]
  $L=1$          &\textbf{20.81}  &\textbf{39.08} &{13.59} &{19.86} &{11.34} &{12.79} &{4.76} &{9.28} \\
  $L=2$          &{20.52}  &{38.41} &\textbf{14.62} &\textbf{21.24} &\textbf{12.52} &\textbf{13.92} &\textbf{5.10} &\textbf{9.62} \\
  $L=3$          &{10.57}  &{33.44} &{12.65} &{19.23} &{6.41} &{7.39} &{3.83} &{7.95} \\
  $L=4$          &{6.32}   &{19.25} &{13.85} &{20.26} &{3.56} &{4.51} &{4.65} &{9.03}\\
\midrule
\multicolumn{9}{|c|}{L$^2$Q-GCN$_{anl}$} \\
\midrule[0.1pt]
  $L=1$          &{20.56}  &{39.51} &{15.82} &{23.04} &{13.11} &{14.50} &{5.52} &{10.31} \\
  $L=2$          &\textbf{22.81}  &\textbf{42.44} &\textbf{16.12} &\textbf{23.62} &\textbf{13.87} &\textbf{15.31} &\textbf{5.74} &\textbf{10.67} \\
  $L=3$          &{21.93}  &{40.85} &{14.60} &{22.22} &{12.90} &{14.52} &{5.35} &{10.14} \\
  $L=4$          &{21.72}  &{40.46} &{14.63} &{22.24} &{11.31} &{13.16} &{4.96} &{9.74} \\
\bottomrule
\end{tabular}}
\end{table}

\begin{table}[bp]
\setlength{\abovecaptionskip}{0cm}
\setlength{\belowcaptionskip}{0cm}
\centering
\caption{Top-20 recommendation (\%) w.r.t different $d$.}
\label{tab:dimension}
\setlength{\tabcolsep}{0.6mm}{
\begin{tabular}{|c | c c | c c | c c | c c |}
\toprule
\multirow{2}*{$L$}  & \multicolumn{2}{c|}{MovieLens} & \multicolumn{2}{c|}{Gowalla} & \multicolumn{2}{c|}{Pinterest} & \multicolumn{2}{c|}{Yelp2018} \\
               ~  & R@20 & N@20 & R@20 & N@20 & R@20 & N@20 & R@20 & N@20\\
\midrule
\midrule
                \multicolumn{9}{|c|}{L$^2$Q-GCN$_{enl}$} \\
\midrule[0.1pt]
  $d=64$           &{16.65}  &{28.23}  &\textbf{14.62}  &\textbf{21.24}  &{9.63}  &{9.69}  &\textbf{5.10}  &\textbf{9.62}    \\
  $d=128$          &{18.02}  &{33.15}  &{14.41}  &{19.91}  &\textbf{12.52}  &\textbf{13.92}  &{4.86}  &{9.63}    \\
  $d=256$          &\textbf{20.52}  &\textbf{38.41}  &{10.39}  &{14.29}  &{10.37}  &{11.62}  &{4.26}  &{8.55}    \\
  $d=512$          &{17.79}  &{31.49}  &{13.23}  &{20.65}  &{10.22}  &{11.38}  &{3.49}  &{7.43}    \\
\midrule
\multicolumn{9}{|c|}{L$^2$Q-GCN$_{anl}$} \\
\midrule[0.1pt]
  $d=64$           &{20.02}  &{38.89}  &{14.29}  &{21.35}  &{11.85}  &{12.76}  &{4.21}  &{8.42}    \\
  $d=128$          &{21.64}  &{40.82}  &{15.38}  &{22.84}  &\textbf{13.87}  &\textbf{15.31}  &{4.94}  &{9.17}    \\
  $d=256$          &\textbf{22.81}  &\textbf{42.44}  &\textbf{16.12}  &\textbf{23.62}  &{12.63}  &{14.20}  &\textbf{5.74}  &\textbf{10.67}    \\
  $d=512$          &{21.23}  &{40.17}  &{15.54}  &{23.11}  &{12.41}  &{14.04}  &{5.28}  &{9.42}    \\
\bottomrule
\end{tabular}}
\end{table}

\begin{algorithm}[bp]
\small
\caption{L$^2$Q-GCN$_{anl}$ algorithm}
\label{alg:anl}
\LinesNumbered  
\KwIn{Interaction graph $\mathcal{G}$; trainable parameters {\footnotesize $\Theta$: $\{\boldsymbol{v}_{u}\}_{u\in\mathcal{U}}$, $\{\boldsymbol{v}_i\}_{i\in\mathcal{I}}$, $\boldsymbol{W}$}; hyper-parameters: {\footnotesize $B$, $d_{end}$, $L$, $\eta$, $\lambda$. } }

\KwOut{Prediction function $\mathcal{F}(u,i|\Theta, \mathcal{G})$} 
$\mathcal{Q}_u \gets \emptyset$, $\mathcal{Q}_i \gets \emptyset$;\\
\While{\rm{L$^2$Q-GCN$_{anl}$ not converge}}{
    \For{$(u,i) \in \mathcal{G}$ \rm{that} $y_{u,i}=1$}{
        \For{$l = 1, \cdots, L$}{
          $\boldsymbol{v}_{u}^{(l)} \gets \sum_{i\in \mathcal{N}(u)} \frac{1}{\sqrt{|\mathcal{N}(u)|\cdot|\mathcal{N}(i)|}}v^{(l-1)}_i$; \\
          $\boldsymbol{v}_{i}^{(l)} \gets \sum_{u\in \mathcal{N}(i)} \frac{1}{\sqrt{|\mathcal{N}(i)|\cdot|\mathcal{N}(u)|}}v^{(l-1)}_u$; \\
          $q_u^{(l)} \gets \big(\boldsymbol{W}^T\boldsymbol{v}^{(l)}_u\big)$; $q_i^{(l)} \gets \big(\boldsymbol{W}^T\boldsymbol{v}^{(l)}_i\big)$; \\
          \If{with quantization}{
              $q_u^{(l)} \gets \sign\big(q_u^{(l)})$; $q_i^{(l)} \gets \sign\big(q_i^{(l)})$; \\
              Setting gradients for backforward propagation; \\
          }
          $\alpha_u^{(l)} \gets \frac{||\boldsymbol{v}_u^{(l)}||_1}{d}$, $\alpha_u^{(l)} \gets \frac{||\boldsymbol{v}_u^{(l)}||_1}{d}$; \\
          Update ($\mathcal{A}_u$, $\mathcal{A}_i$) with $\alpha_u^{(l)}q_u^{(l)}$, $\alpha_i^{(l)}q_i^{(l)}$; \\
        }
      Update ($\mathcal{A}_u$, $\mathcal{A}_i$) with $\alpha_u^{(0)}q_u^{(0)}, \alpha_u^{(0)}q_i^{(0)}$;\\
      $\mathcal{L} \gets $ compute loss and optimize L$^2$Q-GCN$_{anl}$ model;\\ 
    }
}
\KwRet $\mathcal{F}$.\\
\end{algorithm}

\end{document}